\documentclass[conference]{IEEEtran}
\IEEEoverridecommandlockouts
\usepackage{cite}
\usepackage{amsmath,amssymb,amsfonts}
\usepackage{graphicx}
\usepackage{textcomp}
\usepackage{xcolor}
\def\BibTeX{{\rm B\kern-.05em{\sc i\kern-.025em b}\kern-.08em
    T\kern-.1667em\lower.7ex\hbox{E}\kern-.125emX}}
  
\usepackage{amssymb}
\usepackage{gensymb}
\usepackage{graphicx}
\graphicspath{ {images/} }
\usepackage{amsmath}
\usepackage{float}
\usepackage{algorithm}
\usepackage{algpseudocode}
\usepackage{makecell}

\usepackage{geometry}
\begin{document}

\title{ Rate-Splitting Multiple Access for Multi-Antenna Broadcast Channels with Statistical CSIT }

\author{\IEEEauthorblockN{Longfei Yin, Bruno Clerckx and Yijie Mao}
\IEEEauthorblockA{Department of Electrical and Electronic Engineering, Imperial College London, United Kingdom\\
$\mathrm{Email:}\left \{\mathrm{longfei.yin17, b.clerckx, y.mao16} \right \}\mathrm{@imperial.ac.uk}$
\thanks{This work was partially supported by the U.K. Engineering and Physical Sciences Research Council (EPSRC) under EP/R511547/1.}
}
}


\maketitle

\begin{abstract}
Rate-splitting multiple access (RSMA) is a promising technique for downlink multi-antenna communications owning to its capability of enhancing the system performance in a wide range of network loads, user deployments and channel state information at the transmitter (CSIT) inaccuracies.
In this paper, we investigate the achievable rate performance of RSMA in a multi-user multiple-input single-output (MU-MISO) network where only slow-varying statistical channel state information (CSI) is available at the transmitter.
RSMA-based statistical beamforming and the split of the common stream is optimized with the objective of maximizing the minimum user rate subject to a sum power budget of the transmitter. 
Two statistical CSIT scenarios are investigated, namely the Rayleigh fading channels with only spatial correlations known at the transmitter, and the uniform linear array (ULA) deployment with only channel amplitudes and mean of phase known at the transmitter.
Numerical results demonstrate the explicit max min fairness (MMF) rate gain of RSMA over space division multiple access (SDMA) in both scenarios.
Moreover, we demonstrate that RSMA is more robust to the inaccuracy of statistical CSIT. 
\end{abstract}


\section{Introduction}
Multi-user multiple-input multiple-output (MU-MIMO) has attracted significant attention over the past twenty years due to its great potential to boost the system spectral and energy efficiencies. 
Instantaneous channel state information at the transmitter (CSIT) is essential in MU-MIMO to fully achieve its benefits, i.e., to achieve the spatial diversity and multiplexing gain. 
However, acquiring instantaneous CSIT is challenging due to many factors, such as channel estimation errors, limited feedback resources, quantization errors or feedback delay. 
The imperfection of instantaneous CSIT are detrimental to the performance of naive designs that assume perfect CSIT.
In contrast to instantaneous CSI which is changing rapidly and difficult to be obtained at the transmitter, the statistical CSI is commonly slow-varying, and is commonly assumed to be stable over several fading blocks. 
As a consequence, statistical CSI
can be easily and accurately monitored at the transmitter through long-term feedback or covariance extrapolation \cite{9064949}.


For a Rayleigh fading MIMO channel, the statistical CSI is typically captured by the correlation matrix of the channel vector.
Initial works \cite{al2009much,4557521} assume a common transmit correlation matrix for all the users, and naturally leads to a conclusion that transmit correlation can degrade the system performance. 
In \cite{clerckx2008correlated}, the authors further study the scenario when channel statistics vary across users and different users experience different correlation matrices.
With various limited feedback codebooks, transmit correlation is shown to reduce the feedback overhead and quantization error if an appropriate codebook is used.
Spatial correlation is also helpful to the codebook design \cite{clerckx2008mu, huang2010exploiting}.
Spatial correlation-based statistical beamforming with generalized eigenvector solution is studied with the aim of maximizing the ergodic sum rate \cite{raghavan2013statistical} and a lower bound of signal-to-leakage-and-noise-ratio (SLNR) \cite{wang2012statistical} in multi-antenna broadcast channel (BC). 

In addition to spatial correlation, the channel amplitudes usually vary slowly and capture the statistical CSI
especially in some line-of-sight (LOS) channels or atmospheric fading channels.
In \cite{zhang2019robust}, robust beamforming for satellite communications in face of phase perturbations is studied. 
In contrast to typical cellular systems, the channel amplitude which depends on the propagation attenuation is assumed to be constant during the feedback interval.
In \cite{9110855}, slow-varying statistics of channel gain is exploited to solve the difficulty of obtaining instantaneous CSIT in a massive MIMO LEO satellite system.

However, the above works all consider multi-user linear precoding assisted space division multiple access (SDMA). As we know, the major bottleneck of SDMA is the degradation of its degree of freedom (DoF) and spectral efficiency as CSIT becomes worse. 
In contrast to SDMA, rate-splitting multiple access (RSMA) relies on linearly precoded rate-splitting at the transmitter and successive interference cancellation (SIC) at the receivers, which
has been shown to be more spectral efficient \cite{8019852, 8756668,mao2018rate, 7434643,7555358, 7513415, 9257433,8846706} and energy efficient \cite{8846706, 8491100, 9205852} than SDMA in a wide range of network loads, user deployments and channel state information at the transmitter (CSIT) inaccuracies. However, most of the existing works on RSMA assume perfect or imperfect instantaneous CSIT.
There is a lack of investigation on RSMA when the transmitter only knows the statistics of CSI (i.e., the distribution information of user channels). 

In this work, we focus on the statistical CSIT scenarios where the transmitter has only the knowledge of long-term statistics. This contrasts with the existing works on RSMA where instantaneous perfect CSIT \cite{8019852, 8756668, mao2018rate} or imperfect CSIT \cite{7555358, 7513415, 9257433} is assumed.
Statistical beamforming and the message splits of RSMA are jointly designed with
the objective of maximizing the minimum rate.
A weighted minimum mean square error (WMMSE)-based alternating optimization (AO) algorithm is used to transform the initial non-convex problem into convex ones to be iteratively solved.
Two statistical CSIT scenarios are investigated.
In our first scenario considering Rayleigh fading channels with known transmit correlations,
simulation results show the effectiveness of RSMA over SDMA especially when the channels are uncorrelated.
It is also demonstrated 
that high spatial correlation is helpful to the MMF rate performance of both RSMA and SDMA. 
RSMA is more robust to the inaccuracy of correlation coefficient phase.
In the second scenario assuming channel amplitudes and mean of phase known at the transmitter, RSMA shows explicit performance gain over both SDMA and NOMA.

The rest of the paper is organized as follows: the system model and problem formulation of
optimizing MMF rate are described in Section II, followed by the WMMSE and AO algorithm in Section III.
Two scenarios of channel statistics and numerical results are illustrated in Section IV. Section V concludes the paper.

\textit{Notations}: 
Boldface, lowercase and standard letters denote matrices, column vectors, and scalars, respectively. 
$\mathbb{R}$ and $\mathbb{C}$ represent the real and complex domains. 
The real part of a complex number $x$ is given by $\mathcal{R}\left (x  \right )$. 
$\left ( \cdot  \right )^{T}$ and $\left ( \cdot  \right )^{H}$ denote the transpose and the Hermitian transpose, respectively. 
$\left | \cdot  \right |$ and $\left \| \cdot  \right \|$ respectively denote the absolute value and Euclidean norm.

\section{System Model \& Problem Formulation}
We consider a MU-MISO communication network with a base station (BS) equipped with $N_{t}$ transmit antennas serving $K$ single-antenna users.
Let $\mathcal{K}= \left \{ 1,\cdots, K \right \}$ denote all user indices.
The broadcast channel is written as
\begin{equation}
y_{k}=\mathbf{h}_{k}^{H} \mathbf{x} + n_{k},\ k \in \mathcal{K},
\end{equation}
where $y_{k}$ is the received signal at user-$k$, $\mathbf{h}_{k} \in \mathbb{C}^{N_{t} \times 1}$ is the channel vector between the BS and user-$k$. 
$\mathbf{x} \in \mathbb{C}^{N_{t} \times 1}$ is the transmit signal which is subject to a sum power constraint $\mathbb{E}\left \{ \mathbf{x}^{H} \mathbf{x} \right \}\leq P_{t}$.
$n_{k} \in \mathbb{C},\ k  \in \mathcal{K}$ is the complex Gaussian random noise of user-$k$ with zero mean and unit variance, from which the transmit signal-to-noise ratio (SNR) equals to $P_{t}$.

Considering the RSMA strategy at the BS,
the unicast message $W_{k}$ intended for user-$k$ is split into a common sub-message $W_{c,k}$ and a private sub-message $W_{p,k}$.
The common sub-messages $W_{c,1}, \cdots, W_{c,K}$ are jointly encoded into a common stream $s_{c}$, which is intended to be decoded by all users.
The private sub-messages are respectively encoded into $s_{1},\cdots ,s_{K}$.
The vector of symbol streams to be transmitted is $\mathbf{s} = \left [ s_{c},s_{1},\cdots ,s_{K} \right ]^{T} \in \mathbb{C}^{\left ( K+1 \right )\times 1}$, where $\mathbb{E}\big \{ \mathbf{s}\mathbf{s}^{H} \big \}=\mathbf{I}$.
Let $\mathbf{P} = \left [ \mathbf{p}_{c},\mathbf{p}_{c},\cdots ,\mathbf{p}_{K} \right ]\in \mathbb{C}^{\left (N_{t}+1  \right )\times K}$ denotes the linearly precoding matrix, the transmit signal is given as
\begin{equation}
\mathbf{x} = \mathbf{P}\mathbf{s}= \mathbf{p}_{c}s_{c} + \sum_{k=1}^{K}\mathbf{p}_{k}s_{k}.
\end{equation}

Following the decoding process of RSMA, each user sequentially decodes the common stream and the intended private stream to recover its message.
Each user decodes $s_{c}$ at first by treating all private streams as noise.
Then, the common stream is removed from $y_{k}$ through SIC, and each user decodes its private stream by treating the interference from other private streams as noise.
The signal-to-interference-noise ratios (SINRs) of decoding $s_{c}$ and $s_{k}$ at user-$k$ are respectively given as
\begin{equation}
    \gamma _{c,k} = \frac{\left |\mathbf{h}^{H}_{k}\mathbf{p}_{c}  \right |^{2}}{\sum _{i \in \mathcal{K}}\left |\mathbf{h}^{H}_{k}\mathbf{p}_{i}  \right |^{2}+1} ,
 \end{equation}
 \begin{equation}
    \gamma _{k} = \frac{\left |\mathbf{h}^{H}_{k}\mathbf{p}_{k}  \right |^{2}}{\sum _{i \in \mathcal{K}, i \neq k}\left |\mathbf{h}^{H}_{k}\mathbf{p}_{i}  \right |^{2}+1}, \quad \forall k \in \mathcal{K}.
\end{equation}
Therefore, the rates of decoding $s_{c}$ and $s_{k}$ at user-$k$ are $R_{c,k}= \log_{2}\left ( 1+\gamma _{c,k} \right )$ and 
$R_{k}= \log_{2}\left ( 1+\gamma _{k} \right )$.
Since the common stream is shared amongst all users, the rate of the common stream is
$R_{c}= \min_{k \in \mathcal{K}} R_{c,k}$. 
By defining $C_{k}$ as the portion of common rate at user-$k$, we have $R_{c}= \sum _{k \in \mathcal{K}} C_{k}$.
As a consequence, the $k$-th user's rate is denoted by $R_{k,tot}= C_{k}+R_{k}$.


Next, we investigate the ergodic rate of each user with statistical CSI at the transmitter.
Ergodic rate is known as a long-term measure which captures the expected performance over the known channel distribution information. 
With given statistical CSIT,
we can construct the set of i.i.d channel samples 
$\mathbb{H}^{\left ( S \right )} =\left \{ \mathbf{H}^{\left ( s \right )}, s \in \mathfrak{S}=\left \{ 1, \cdots, S \right \} \right \}$.
When the sample size $S\rightarrow \infty$, based on the strong law of large numbers, the ergodic common and private rate of each user are approximated by the following sample average functions (SAFs).
\begin{align}
\overline{R}_{c,k}& = \lim_{S\rightarrow \infty} \overline{R}_{c,k}^{\left (S  \right )} =\lim_{S\rightarrow \infty} \frac{1}{S}\sum_{s=1}^{S} R_{c,k}\big ( \mathbf{H}^{\left (s  \right )} \big ),
\\
 \overline{R}_{k}& = \lim_{S\rightarrow \infty}\overline{R}_{k}^{\left (S  \right )} =\lim_{S\rightarrow \infty} \frac{1}{S}\sum_{s=1}^{S} R_{k}\big ( \mathbf{H}^{\left (s  \right )} \big ),
 \end{align}
 where $R_{c,k}\left ( \mathbf{H}^{\left (s  \right )} \right )$ and $R_{k}\left ( \mathbf{H}^{\left (s  \right )} \right )$ are rates correspond to the $s$-th channel sample. 
 All samples are available at the transmitter so as to approximate the ergodic rate.
 With the given statistical CSIT, we formulate the optimization problem $\overline{\mathcal{R}}$ to achieve max min fairness (MMF) amongst all users subject to a sum transmit power constraint, which is given by
\begin{align}
\overline{\mathcal{R}}: \quad
\max _{\mathbf{\overline{c}},\mathbf{P}} &\min_{k\in \mathcal{K}} \big ( \overline{C}_{k} +  \overline{R}_{k}^{\left ( S \right )}\big )
\\
s.t. \quad
&\overline{R}_{c,k}^{\left ( S \right )}\geq \sum_{k=1}^{K} \overline{C}_{k},\quad  \forall k \in \mathcal{K}
\\
&\overline{C}_{k} \geq 0, \quad  \forall k \in \mathcal{K}
\\
&\left \| \mathbf{p}_{c} \right \|^{2}+\sum _{k=1}^{K}\left \| \mathbf{p}_{k} \right \|^{2}\leq P_{t},
\end{align}
where
$\mathbf{\overline{c}} = \big [\overline{C}_{1},\cdots ,\overline{C}_{K}  \big ]$ is the vector of ergodic common rate allocation among users, which is optimized together with the precoding matrix $\mathbf{P}$.
Note that $\mathbf{P}$ is fixed over each $R_{c,k}\big ( \mathbf{H}^{\left (s  \right )} \big )$ and $R_{k}\big ( \mathbf{H}^{\left (s  \right )} \big )$.
Constraint (8) ensures that the common stream $s_{c}$ is decoded by each user.
Constraint (9) ensures that each entry of $\mathbf{\overline{c}}$ is non-negative. 
$P_{t}$ is the sum transmit power budget, and (10) implies the sum power constraint.

\section{WMMSE Optimization Framework}
The optimization problem described above is non-convex.
To solve the RSMA-based problem, we follow the method in \cite{7555358} to reformulate the original non-convex problem into an equivalent WWMSE form, and solve the reformulated problem based on an AO algorithm.
At user-$k$, the stream which is decoded first is $s_{c}$.
The estimate of $s_{c}$ is given by $\widehat{s}_{c,k}=g_{c,k}y_{k}$, where $g_{c,k}$ represents the equalizer. 
After the common stream is decoded and removed by SIC, the estimate of desired $s_{k}$ is $\widehat{s}_{ k }=g_{k}\left (y_{k} - \mathbf{h}_{k}^{H} \mathbf{p}_{c}s_{c}\right )$.
The common and private mean square errors (MSEs) are respectively
\begin{align}
\varepsilon _{c,k}& =\left | g_{c,k}\right |^{2} T_{c,k}-2\mathcal{R}\big \{ g_{c,k}\mathbf{h}_{k}^{H} \mathbf{p}_{c} \big \} +1,
\\
\varepsilon _{k}&=\left | g_{k}\right |^{2} T_{k}-2\mathcal{R}\big \{ g_{k}\mathbf{h}_{k}^{H} \mathbf{p}_{k} \big \}+1,
\end{align}
where we have $T_{c,k}=\left|\mathbf{h}_{k}^{H}\mathbf{p}_{c}  \right |^{2}+T_{k}$ and 
$ T_{k}= \left |\mathbf{h}_{k}^{H}\mathbf{p}_{k }  \right |^{2}+\sum _{j=1,j\neq k}^{K}\left |\mathbf{h}_{k}^{H}\mathbf{p}_{j}  \right |^{2}+ 1$.
The minimum MSE (MMSE) equalizers are given by
\begin{align}
    g_{c,k}^{MMSE}=\mathbf{p}_{c}^{H}\mathbf{h}_{k}T_{c,k}^{-1} \quad 
   \mathrm{and} \quad
    g_{k}^{MMSE}=\mathbf{p}_{k}^{H}\mathbf{h}_{k}T_{k}^{-1}.
\end{align}
By substituting (13) into (11) and (12), the MMSEs are 
\begin{align}
\varepsilon _{c,k}^{MMSE} &= \min_{g_{c,k}} \varepsilon _{c,k} = T_{c,k}^{-1}I_{c,k}   ,
\\
\varepsilon _{k}^{MMSE} &= \min_{g_{k}} \varepsilon _{k} = T_{k}^{-1}I_{k} ,  
\end{align}
where we define $I_{c,k}=T_{c,k}$ as the interference portion in $T_{c,k}$ and 
$I_{k}=T_{k}-\left |\mathbf{h}_{k}^{H}\mathbf{p}_{k}  \right |^{2}$ as the interference portion in $T_{k}$. 
Thus, the SINRs can be written in the form of MMSEs, i.e., 
$\gamma _{c,k} = \big ( 1/\varepsilon _{c,k}^{MMSE} \big )-1$ and $\gamma _{k} = \left ( 1/\varepsilon _{k}^{MMSE} \right )-1$.
As a result, the rates are expressed by $R_{c,k}=-\log_{2}\big (\varepsilon _{c,k}^{MMSE}   \big )$ and $R_{k}=-\log_{2}\left (\varepsilon _{k}^{MMSE}   \right )$. 

Furthermore, the common and private weighted MSEs (WMSEs) of user-$k$ are given by
\begin{equation}
\xi _{c,k}= u _{c,k} \varepsilon _{c,k}-\log _{2} u _{c,k}
\
\mathrm{and}
\
\xi _{k}= u _{k} \varepsilon _{k}-\log _{2} u _{k},
\end{equation}
where $u_{c,k}$ and $u_{k}$ are the weights associated with MSEs. 
To obtain the minimum WMSEs (WMMSEs) over both equalizers and weights,
we substitute the MMSE equalizers to (16) and let $\frac{\partial \xi _{c,k}\left ( g _{c,k}^{MMSE} \right )}{\partial u_{c,k}}=0$ and $\frac{\partial \xi _{k}\left ( g _{k}^{MMSE} \right ) }{\partial u_{k}}=0$. Thereby, the MMSE weights are obtained as 
\begin{equation}
    u _{c,k}^{MMSE}=\big ( \varepsilon  _{c,k}^{MMSE} \big )^{-1} 
\
\mathrm{and}
\
u _{k}^{MMSE}=\left ( \varepsilon  _{k}^{MMSE} \right )^{-1}.
\end{equation}
We substitute (13) and (17) into (16), and the rate-WMMSE relationship is obtained.
\begin{align}
\xi_{c,k}^{MMSE} &= \min_{g_{c,k},u_{c,k}}\xi _{c,k}= 1+ \log_{2}\varepsilon _{c,k}^{MMSE} = 1- R_{c,k},
\\
\xi_{c,k}^{MMSE} &= \min_{g_{c,k},u_{c,k}}\xi _{c,k}
=1+ \log_{2}\varepsilon _{k}^{MMSE}= 1- R_{k}.
\end{align}

By taking the expectation over the known channel distribution and applying SAFs for approximation \cite{7555358}, the SAF version of the rate-WMMSE relationship writes as
\begin{align}
\overline{\xi}_{c,k}^{MMSE\left (S \right )} &= \min_{\mathbf{g}_{c,k},\mathbf{u}_{c,k}}\overline{\xi} _{c,k}^{\left ( S \right )}= 1- \overline{R}_{c,k}^{\left ( S \right )}   ,
\\
\overline{\xi}_{k}^{MMSE\left (S  \right )} &= \min_{\mathbf{g}_{k},\mathbf{u}_{k}}\overline{\xi} _{k}^{\left ( S \right )}= 1- \overline{R}_{k}^{\left ( S \right )},
\end{align}
where $\overline{\xi}_{c,k}^{MMSE\left (S \right )}=\frac{1}{S}\sum ^{S}_{s=1}\xi_{c,k}^{MMSE\left (s  \right )}$ and $\overline{\xi}_{k}^{MMSE\left (S  \right )}=\frac{1}{S}\sum ^{S}_{s=1}\xi_{k}^{MMSE\left (s \right )}$ represent SAF approximations of the ergodic WMMSEs when $S\rightarrow \infty$.
$\xi_{c,k}^{MMSE\left (s  \right )}$ and $\xi_{k}^{MMSE\left (s \right )}$ are associated with the $s$-th sample in $\mathbb{H}^{\left ( S \right )}$. 
The sets of MMSE equalizers are defined as $\mathbf{g}_{c,k}^{MMSE}=\big \{ g_{c,k}^{MMSE\left ( s \right )} \mid s \in \mathfrak{S}\big \}$ and $\mathbf{g}_{k}^{MMSE}=\big\{ g_{k}^{MMSE\left ( s \right )} \mid s \in \mathfrak{S}\big \}$. 
The sets of MMSE weights are  $\mathbf{u}_{c,k}^{MMSE}=\big \{ u_{c,k}^{MMSE\left ( s \right )} \mid s \in \mathfrak{S}\big \}$ and $\mathbf{u}_{k}^{MMSE}=\big \{ u_{k}^{MMSE\left ( s \right )} \mid s \in \mathfrak{S}\big \}$. 
For compactness, we define the composite MMSE equalizer and weight of all the users as $\mathbf{G}^{MMSE} = \big \{ \mathbf{g}_{c,k}^{MMSE},\mathbf{g}_{k}^{MMSE}\mid k \in \mathcal{K} \big \}$ and $\mathbf{U}^{MMSE}= \big \{ \mathbf{u}_{c,k}^{MMSE},\mathbf{u}_{k}^{MMSE}\mid k \in \mathcal{K} \big \}$.
Motivated by the rate-WMMSE relationship, the reformulated equivalent WMMSE problem is written as
\begin{align}
\overline{\mathcal{W}}: \quad
&\max _{\mathbf{\overline{c}},\mathbf{P},\mathbf{G},\mathbf{U},\overline{r}_{g}}\overline{r}_{g}
\\
s.t. \quad
&\overline{C}_{k}+\big ( 1- \overline{\xi}_{k}^{\left (S  \right )} \big )\geq \overline{r}_{g},\quad  \forall k \in \mathcal{K}
\\
&1-\overline{\xi}_{c,k}^{\left ( S \right )}\geq \sum_{k=1}^{K} \overline{C}_{k},\quad  \forall k \in \mathcal{K}
\\
&\overline{C}_{k} \geq 0, \quad  \forall k \in \mathcal{K}
\\
&\left \| \mathbf{p}_{c} \right \|^{2}+\sum _{k=1}^{K}\left \| \mathbf{p}_{k} \right \|^{2}\leq P_{t},
\end{align}
where $\overline{r}_{g}$ is an auxiliary variable.
For any stationary point of $\overline{\mathcal{W}}$ given by
$\left (\mathbf{P}^{*},\mathbf{G}^{*},\mathbf{U}^{*}, \overline{r}_{g}^{*}, \overline{\mathbf{c}}^{*} \right )$, there exists a stationary point of $\overline{\mathcal{R}}$ given by
$\left (\mathbf{P}^{*},  \overline{\mathbf{c}}^{*} \right )$. 
Although $\overline{\mathcal{W}}$ is still non-convex with respect to the joint set of optimization variables, it is block-wise convex, e.g., the problem is convex in $\mathbf{P}$ when assuming $\mathbf{G}$, $\mathbf{U}$ and fixed. 
Note that the MMSE solutions of $\left ( \mathbf{G},\mathbf{U} \right )$ associated with the rate-WMMSE relationship are optimum for $\overline{\mathcal{W}}$.
Therefore, an AO algorithm described in Algorithm 1 is utilized to solve $\overline{\mathcal{W}}$.

In the $n$-th iteration of the algorithm, based on the precoding matrix $\mathbf{P}^{\left [ n-1 \right ]}$ obtained from the previous iteration, the equalizers and weights are updated by closed form MMSE solutions $\mathbf{G}^{MMSE}\big (\mathbf{P}^{\left [ n-1 \right ]}  \big)$ and $\mathbf{U}^{MMSE}\big (\mathbf{P}^{\left [ n-1 \right ]}  \big)$. Then, with the updated $\mathbf{G}$ and $\mathbf{U}$, 
we can write the SAF expressions of average WMMSE as
\begin{align}
    \overline{\xi}_{c,k}^{\left (S  \right )} &= 
\mathbf{p}_{c}^{H} \overline{\Psi }_{c,k}^{\left ( S \right )} \mathbf{p}_{c} 
 +\sum _{i=1}^{K}\mathbf{p}_{i}^{H} \overline{\Psi }_{c,k}^{\left ( S \right )} \mathbf{p}_{i}  + \overline{t}_{c,k}^{\left ( S \right )} -2 \mathcal{R}\big \{ \overline{\mathbf{f} }_{c,k}^{\left ( S \right )H} \mathbf{p}_{c}\big \}  \notag \\&+ \overline{u}_{c,k}^{\left (S  \right )}  
 -\overline{v}_{c,k}^{\left (S  \right )},
 \\
    \overline{\xi}_{k}^{\left (S  \right )} &= \sum _{i=1}^{K}\mathbf{p}_{i}^{H} \overline{\Psi }_{k}^{\left ( S \right )} \mathbf{p}_{i} + \overline{t}_{k}^{\left ( S \right )}  -2 \mathcal{R}\big \{ \overline{\mathbf{f} }_{k}^{\left ( S \right )H} \mathbf{p}_{k}\big \}    +\overline{u}_{k}^{\left (S  \right )}-\overline{v}_{k}^{\left (S  \right )},
\end{align}
where $\overline{t}_{c,k}^{\left ( S \right )},\  \overline{t}_{k}^{\left ( S\right )},\  \overline{\Psi }_{c,k}^{\left ( S \right )},\  \overline{\Psi }_{k}^{\left ( S \right )},\  \overline{\mathbf{f} }_{c,k}^{\left ( S \right )},\  \overline{\mathbf{f} }_{k}^{\left ( S \right )},\  \overline{v }_{c,k}^{\left ( S \right )},\  \overline{v }_{k}^{\left ( S \right )}$ 
are SAFs obtained according to the updated $\mathbf{G}$ and $\mathbf{U}$.
Details can be found in \cite{7555358}.
By substituting equation (27) and (28) into $\overline{\mathcal{W}}$ and removing $\mathbf{G}$ and $\mathbf{U}$ from optimization variables,
$\overline{\mathcal{W}}$ becomes convex and the precoding matrix $\mathbf{P}$ is therefore optimized.
Updating $\mathbf{G}$, $\mathbf{U}$ and updating $\mathbf{P}$ are repeated
alternatively until convergence. $\varepsilon = 10^{-4}$ is the tolerance of the algorithm, which determines the accuracy of the optimization.

\begin{algorithm}
\caption{Alternating Optimization}\label{A}
\begin{algorithmic}[1]
\State \textbf{Initialize}: $n\leftarrow 0,\  \overline{\mathcal{W}}^{\left [ n \right ]}\leftarrow 0, \mathbf{P}$.
\Repeat
\State $n\leftarrow n+1$, $\mathbf{P}^{\left [ n-1 \right ]}\leftarrow \mathbf{P}$.
\State $\mathbf{G}\leftarrow \mathbf{G}^{MMSE}\left ( \mathbf{P} ^{\left [ n-1 \right ]}\right )$, $\mathbf{U}\leftarrow \mathbf{G}^{MMSE}\left ( \mathbf{U} ^{\left [ n-1 \right ]}\right )$.
\State update $\overline{t}_{c,k}^{\left ( S \right )},  \overline{t}_{k}^{\left ( S\right )},  \overline{\Psi }_{c,k}^{\left ( S \right )},  \overline{\Psi }_{k}^{\left ( S \right )},  \overline{\mathbf{f} }_{c,k}^{\left ( S \right )},  \overline{\mathbf{f} }_{k}^{\left ( S \right )}$, $\overline{v }_{c,k}^{\left ( S \right )},  \overline{v }_{k}^{\left ( S \right )}$, 

for all $k\in \mathcal{K}$.
\State $\mathbf{P}\leftarrow \arg \overline{\mathcal{W}}^{\left [ n \right ]}$.
\Until{$\left |\overline{\mathcal{W}}^{\left [ n \right ]}-\overline{\mathcal{W}}^{\left [ n-1 \right ]}  \right |< \varepsilon $.}
\end{algorithmic}
\end{algorithm}

\section{Simulation Results}

In this section, we evaluate the simulation results based on the above algorithm.
Two different statistical CSIT scenarios are investigated.
The former scenario is commonly used for cellular systems, while the latter is more suited for non-terrestrial systems.

In the first scenario, we consider a Rayleigh fading channel model where the channel statistics described by the spatial correlation matrices are known at the transmitter. 
The channel vector between the BS and user-$k$ is modeled as $\mathbf{h}_{k}=\mathbf{R}_{t,k}^{1/2}\mathbf{h}_{w,k}$, where
$\mathbf{h}_{w,k}, \ k \in \mathcal{K}$ are i.i.d channels with entries drawn from $\mathcal{C}\mathcal{N}\left ( 0,1 \right )$ \cite{clerckx2008correlated}.
By taking a 4-antenna transmitter as an example, user-$k$'s transmit correlation matrix $\mathbf{R}_{t,k}$ is given by 
\begin{equation}
\mathbf{R}_{t,k} = \mathbb{E}\left \{ \mathbf{h} \mathbf{h}^{H} \right \} = \begin{bmatrix}
1 & t_{k} &t_{k}^{2}  &t_{k}^{3} \\ 
t_{k}^{*} & 1 & t_{k} &t_{k}^{2}\\ 
t_{k}^{*2} & t_{k}^{*} & 1 & t_{k}\\ 
t_{k}^{*3} & t_{k}^{*2} & t_{k}^{*}  & 1 
\end{bmatrix},
\end{equation}
where $t_{k}$ is the transmit correlation coefficient for user-$k$.
The eigenvalue decomposition of $\mathbf{R}_{t,k}$ writes as
\begin{equation}
    \mathbf{R}_{t,k} = \mathbf{U}_{t,k} \mathbf{\Lambda }_{t} \mathbf{U}_{t,k} ^{H},
\end{equation}
where $\mathbf{\Lambda }_{t}$ is a diagonal matrix containing eigenvalues $\lambda_{1},\cdots ,\lambda_{\mathrm{rank} \left (\mathbf{\Lambda }_{t}  \right )}$ ordered decreasingly, i.e., $\lambda_{1}\geq \lambda_{2}\geq\cdots \geq \lambda_{\mathrm{rank} \left (\mathbf{\Lambda }_{t}  \right )}$. 
The eigenvectors in $\mathbf{U}_{t,k}$ represent dominant transmit directions.
Specifically, the eigenvalues are functions of the magnitude of $t_{k}$, while the eigenvectors are functions of the phase of $t_{k}$. 
To evaluate the influence of spatial correlation on the system performance, all users are assumed to have the same magnitude $\left | t_{k} \right | =\left |t  \right |, \forall k \in \mathcal{K}$ by ignoring the user indices. 
This is motivated by the fact that $\left |t  \right |$ is a function of the BS inter-element spacing \cite{clerckx2008correlated}. 
However, the phases $ \frac{t_{k}}{\left | t \right |}$ are independent from each other and uniformly distributed over $\left [ - \pi, \pi\right ]$.
With such assumption, the matrix $\mathbf{U}_{t,k}$ is independent from one user to another.
The eigenvalues in $\mathbf{\Lambda }_{t}$ are equal for all users. 
When $\left |t  \right | = 0$, each user's channel is spatially uncorrelated.
$\mathbf{\Lambda }_{t} = \mathbf{I}$, and $\mathrm{rank} \left (\mathbf{\Lambda }_{t}  \right ) = N_{t}$. Otherwise, $\left |t  \right | = 1$ indicates fully correlated channels with $\mathrm{rank} \left (\mathbf{\Lambda }_{t}  \right ) = 1$.

In the second scenario, we consider a LOS
uniform linear array (ULA) deployment where the channel phases change more rapidly than the amplitudes.
When $N_{t} = 4$, the channel vector of user-$k$ can be written as
\begin{equation}
    \mathbf{h}_{k}=\beta _{k}\times \left [1 , e^{j \varphi_{k}} ,e^{j 2\varphi_{k}},e^{j 3\varphi_{k}} \right ]^{H}.
\end{equation}
For each user, we assume the channel amplitude $\beta _{k}$ is known. 
The phase $\varphi_{k}$ is uniformly distributed over $\left [\varphi_{k}^{min},\ \varphi_{k}^{max}  \right ]$, and has a mean $\overline{\varphi}_{k}$ which is user dependent.
All users are assumed to have the same range of phase $\mathrm{r} = \varphi_{k}^{max} - \varphi_{k}^{min}$.



In Fig. 1, we consider the Rayleigh fading channels with known transmit correlation matrices. 
Following the assumption in the system model, all users have the same magnitude of transmit correlation coefficient, i.e.,
$\left |t_{k}  \right |= \left |t  \right | = 1,\ 0.9,\ 0.6$ or $0$ during the simulations.
The correlation coefficient phases of all users
$\frac{t_{k}}{\left | t \right |}$
 are independent from each other and randomly drawn from $\left [ -\pi, \pi\right ]$.
 Therefore, we obtain the set of independent transmit correlation matrices
 $\mathcal{R}_{t} = \left \{ \mathbf{R}_{t,1},\cdots ,\mathbf{R}_{t,K} \right \}$.
 The set of channel samples $\mathbb{H}^{\left ( S \right )}$
 is constructed according to the given channel statistics and is available at the transmitter to approximate ergodic rates by the SAFs. The sample size $S = 1000$ is used.
 It should be noted that all solutions in Fig. 1 are obtained by averaging over $100$ different $\mathcal{R}_{t} $ based on fixed $\left |t  \right | $ and $100$ randomly selected $\frac{t_{k}}{\left | t \right |}$ 
 to evaluate the MMF rate performance. 
Here, we consider $N_{t} = 8$ transmit antennas and $K = 6$ users in the MU-MISO system.
The results of perfect instantaneous CSIT are also provided.
SDMA is used as the benchmark.
It is a special case of RSMA by turning off the common stream. 
Readers are referred to \cite{mao2018rate} for the detailed comparison between SDMA and RSMA. 
NOMA is not adopted in this scenario due to the fact that multi-antenna NOMA is not suited for general user deployments and results in a waste of DoF and therefore rate loss.
It is suited when users are sufficiently aligned with each other and exhibit a disparity of channel strengths.
NOMA also leads to higher complexity with multi-layer SIC at the users.
It is observed that 
the performance with perfect CSIT is better than the statistical CSIT scenarios for both RSMA and SDMA.
Since the system is underloaded, 
both transmit schemes achieve equal MMF-DoF, which is $1$ under perfect CSIT. 
For statistical CSIT, the MMF rate performance improves as $\left |t  \right |$ grows. 
When $\left |t  \right | = 1$, all the channels are spatially correlated. 
Only one eigenvalue exists in $\mathbf{\Lambda }_{t}$.
The maximum eigenvector of each $\mathbf{U}_{t,k}$ indicates exactly the space channel direction. 
As a consequence, the beamforming optimization based on statistical CSIT when $\left |t  \right | = 1$ is very accurate.
Spatially correlated channels are indeed beneficial to the statistical beamforming design and MMF rate performance for both RSMA and SDMA.
RSMA shows better rate performance compared with SDMA due to its more flexible architecture.
Otherwise, when $\left |t  \right | \neq 1$, the eigenvectors of $\mathbf{U}_{t,k}$ represent different dominant directions, and there exists more than one eigenvalues in $\mathbf{\Lambda }_{t}$.
The maximum eigenvalue decreases as $\left |t  \right |$ drops. Finally, when $\left |t  \right | = 0$, all channels are spatially uncorrelated. $\mathbf{\Lambda }_{t}$ becomes an identity matrix and the MMF rate mostly degrades.
As we can see, the DoFs of RSMA curves are around $\frac{1}{K}$.
For SDMA, all DoFs reduce to $0$, i.e., the ceiling effect is observed.
The DoF performance coincides with the results of non-scaling CSIT with a fixed number of feedback bits in \cite{9257433}.
Above all, 
RSMA can always provide MMF rate gains over SDMA, and also provide DoF gains when the channels are not spatially correlated.

\begin{figure}
\vspace{-1em}
    \centering
    \includegraphics[width=0.42\textwidth]{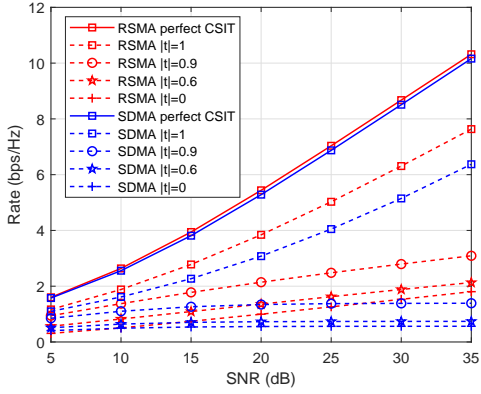}
    \caption{MMF rate performance. $N_{t}=8$ antennas, $K=6$ users.
    The magnitude of the transmit correlation coefficient $\left |t  \right | = 1,\ 0.9,\ 0.6,\ 0$.  The phase  $t_{k}/\left |t  \right |$
 is uniformly distributed over $\left [ -\pi, \pi\right ]$.}
    \label{fig:Fig1}
\end{figure}

\begin{figure}
\vspace{-1em}
    \centering
    \includegraphics[width=0.42\textwidth]{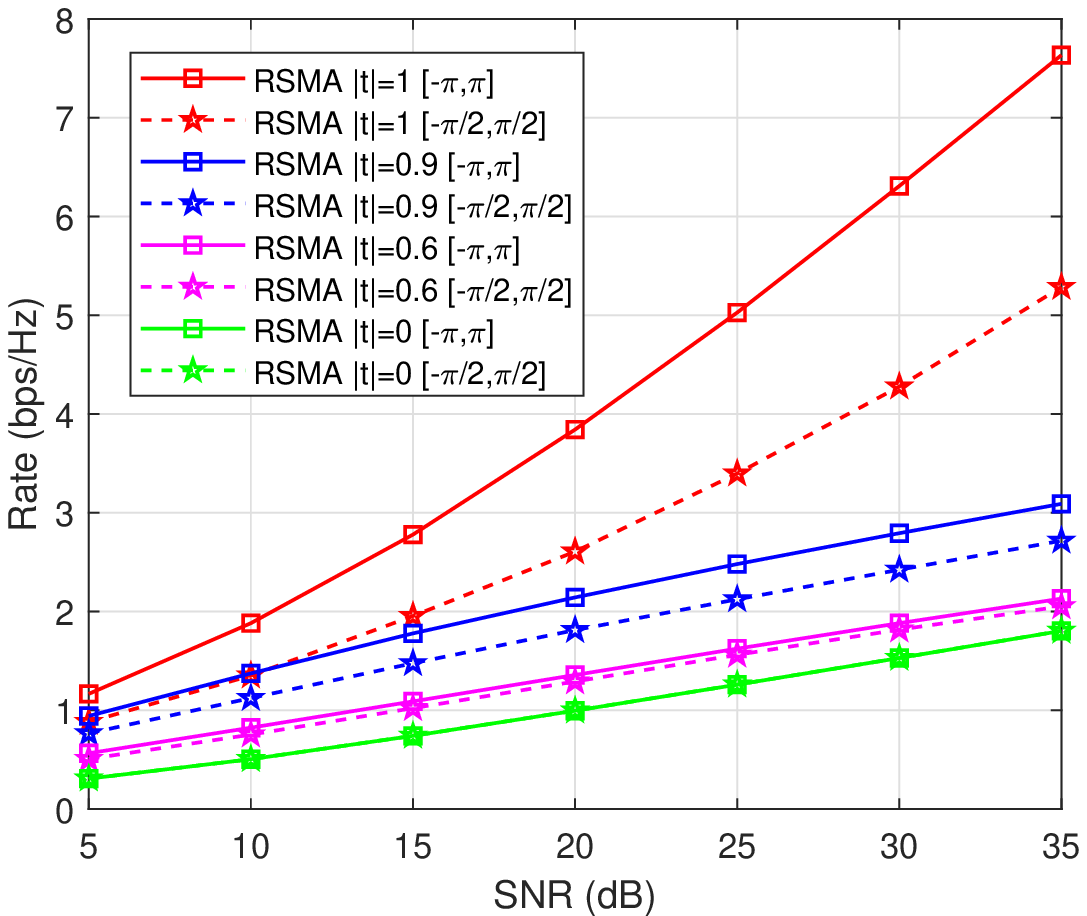}
    \caption{MMF rate performance. $N_{t}=8$ antennas, $K=6$ users.
    The phase of the transmit correlation coefficient $t_{k}/\left |t  \right |$
 is uniformly distributed over $\left [ -\pi, \pi\right ]$ or $\left [ -\frac{\pi}{2}, \frac{\pi}{2}\right ]$.
    }
    \label{fig:Fig21}
\end{figure}

\begin{figure}
\vspace{-1em}
    \centering
    \includegraphics[width=0.42\textwidth]{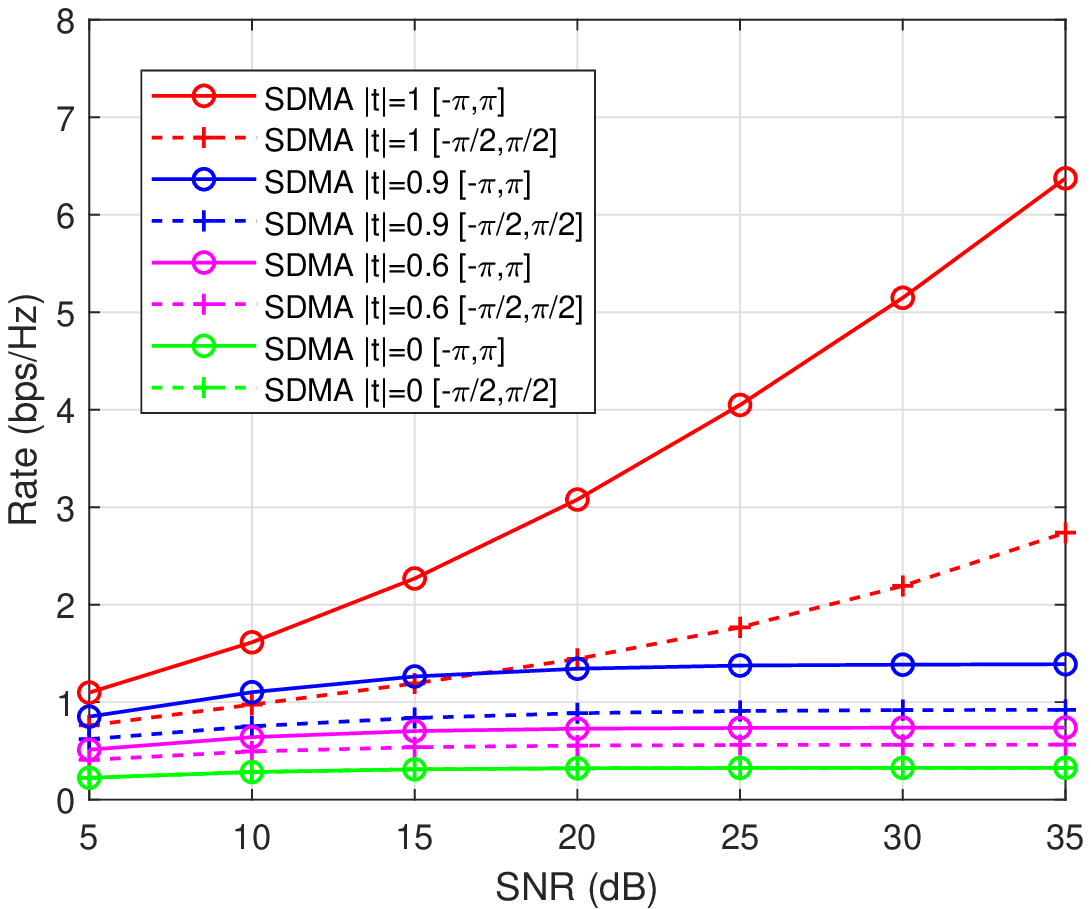}
    \caption{MMF rate performance. $N_{t}=8$ antennas, $K=6$ users.
    The phase of the transmit correlation coefficient $t_{k}/\left |t  \right |$
 is uniformly distributed over $\left [ -\pi, \pi\right ]$ or $\left [ -\frac{\pi}{2}, \frac{\pi}{2}\right ]$.
    }
    \label{fig:Fig22}
\end{figure}

We further study the influence of the phase of correlation coefficient.
In Fig. 2 and Fig. 3, we consider the scenario where all $\frac{t_{k}}{\left | t \right |}$ are independent from each other and uniformly generated over
$\left [ -\frac{\pi}{2},\frac{\pi}{2}\right ]$. 
The simulation results are still obtained by averaging over $100$ $\mathcal{R}_{t} $. 
Compared with the results of $\frac{t_{k}}{\left | t \right |}$ drawing from $\left [ - \pi, \pi\right ]$,
it is found that limited phase distribution is very detrimental to the MMF performance in the presence of high transmit correlations.
When $\left |t  \right | = 1$, each space channel direction is determined by the corresponding maximum eigenvector. 
Since the eigenvector is a function of the correlation coefficient phase $\frac{t_{k}}{\left | t \right |}$, reducing the range of the phase distribution from $\left [ - \pi, \pi\right ]$ to $\left [ -\frac{\pi}{2},\frac{\pi}{2}\right ]$ for all $k \in \mathcal{K}$ increases the user correlation significantly, and therefore it restricts the spectrum efficiency.
By comparing the red curves in Fig. 2 for RSMA with those in Fig. 3 for SDMA, we observe that the MMF rate of SDMA drops significantly as the range of the phase for the transmit correlation coefficient decreases from $\left [ - \pi, \pi\right ]$ to $\left [ -\frac{\pi}{2},\frac{\pi}{2}\right ]$.
In comparison, RSMA is more robust to the channel nonorthogonality, which coincides with the results obtained in \cite{mao2018rate}.

All results above assume accurate statistical CSIT, i.e., the transmit correlation matrices are perfectly known at the transmitter. 
Next, we investigate the influence of the statistical CSIT inaccuracy. Specifically, we consider a special case where $\frac{t_{k}}{\left |t  \right |}$ is imperfectly known at the transmitter. 
For example, the inaccuracy range $\frac{\pi}{4}$ means that the transmitter only knows that $\frac{t_{k}}{\left |t  \right |}$ is uniformly distributed over $\big [\frac{t_{k}}{\left |t  \right |} - \frac{\pi}{8},\frac{t_{k}}{\left |t  \right |} + \frac{\pi}{8}  \big ]$, rather than the exact value.
With such assumption, the distribution of inaccurate $\frac{t_{k}}{\left |t  \right |}$ should also be considered when we construct $\mathbb{H}^{\left ( S \right )}$.
From Fig. 4, the MMF rate gain of RSMA over SDMA is still obvious when the information of correlation coefficient phases at the transmitter is inaccurate. 
Hence, RSMA is more robust to the statistical CSIT inaccuracy than SDMA.
Compared with accurate statistical CSIT scenarios, when $\left |t  \right |=1$,
the MMF rate performance of both RSMA and SDMA degrades a lot.
The reason is that the benefit brought by high spatially correlated channels relies significantly on the accuracy of correlation coefficient phase information.
When the channels are not spatially correlated, taking $\left |t  \right |=0.6$ as an example, 
the influence of $\frac{t_{k}}{\left |t  \right |}$ inaccuracy is very tiny.

\begin{figure}
\vspace{-1em}
    \centering
    \includegraphics[width=0.42\textwidth]{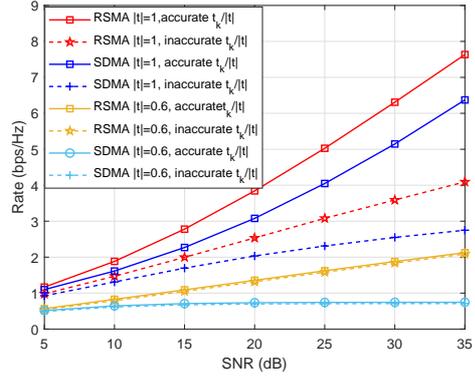}
    \caption{MMF rate performance. $N_{t}=8$ antennas, $K=6$ users.
    The phase of the transmit correlation coefficient $t_{k}/\left |t  \right |$
 is uniformly distributed over $\left [ -\pi, \pi\right ]$. The range of inaccurate $t_{k}/\left |t \right |$ is $\frac{\pi}{4}$.
    }
    \label{fig:Fig4}
\end{figure}

\begin{figure}
\vspace{-1em}
    \centering
    \includegraphics[width=0.42\textwidth]{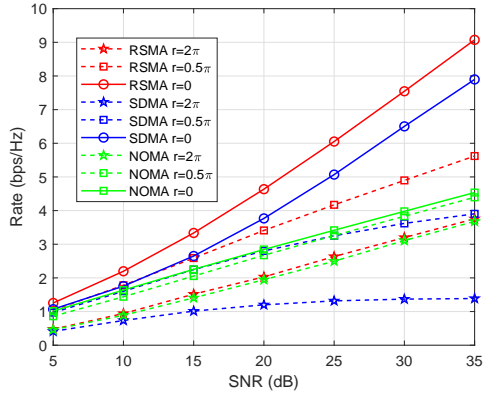}
    \caption{MMF rate performance. $N_{t}=4$ antennas, $K=3$ users.
    The range of the phase $\mathrm{r}= 2 \pi,\ 0.5 \pi, \ 0$.
    }
    \label{fig:Fig3}
\end{figure}

Fig. 5 illustrates the results of the second statistical CSIT scenario with channel amplitudes and mean of phase known at the transmitter.
Here, we consider $N_{t} = 4$ antennas and $K = 3$ users.
The channel amplitudes are assumed to be $\beta_{1} = 1$, $\beta_{2} = 0.8$ and $\beta_{3} = 0.5$. 
The mean of phase $\overline{\varphi}_{k}$ for each user is generated randomly.
Different phase ranges $\mathrm{r}$ are investigated.
With given statistical CSIT, the set of channel samples $\mathbb{H}^{\left ( S \right )}$ is constructed to approximate ergodic rates. 
$S = 1000$ samples are used.
Note that the MMF rate performance in Fig. 5 is evaluated by averaging over $100$ random mean of phase selections. 
Both SDMA and NOMA are adopted as the baseline in this scenario with a disparity of channel strengths.
We observe that RSMA always outperforms SDMA and NOMA. 
The setting of $\mathrm{r}= 0$ is known as the extreme case with perfect instantaneous CSIT.
RSMA and SDMA have the same DoF, which is $1$. 
RSMA performs slightly better than SDMA in the rate sense owning to its more flexible architecture. 
NOMA performs the worst as one of the three users is required to decode all the streams and thus the DoF is sacrificed to $\frac{1}{K}$.
When the range of unknown phase increases to $\mathrm{r}= 2\pi$,
the DoF of RSMA and SDMA decreases to $\frac{1}{K}$ and $0$, respectively.
Both MMF rate gain and DoF gain of RSMA over SDMA become very explicit.
However, the DoF of NOMA remains unchanged and it performs very close to RSMA.
We conclude that NOMA is more suited to the wide range of phase scenario than SDMA. 
RSMA softly bridges NOMA and SDMA through partially decoding interference and partially treating interference as noise, and outperforms both schemes in any $\mathrm{r}$ setting.

\section{Conclusion}
In this paper, we investigate the application of RSMA for multi-antenna BC with statistical CSIT. 
Statistical beamforming is obtained by optimizing MMF rate subject to a sum power constraint at the transmitter.
 Two statistical CSIT scenarios are investigated.
 In the first scenario of Rayleigh fading channels with only spatial correlations known at the transmitter,
 simulation results show that transmit correlated fading is beneficial to the statistical beamforming design and MMF rate performance for both RSMA and SDMA. 
 RSMA achieves appealing MMF rate gain over the benchmark SDMA.
 DoF gain appears when the channels are not spatially correlated.
 RSMA is also demonstrated to be more robust to statistical CSIT inaccuracy.
 In the second scenario considering uniform linear array (ULA) deployment with distinct channel amplitudes and mean of phase known at the transmitter,
the MMF rate gain of RSMA over SDMA and NOMA is verified.

\bibliographystyle{IEEEtran}
\bibliography{ref}

\begin{thebibliography}{10}
\providecommand{\url}[1]{#1}
\csname url@samestyle\endcsname
\providecommand{\newblock}{\relax}
\providecommand{\bibinfo}[2]{#2}
\providecommand{\BIBentrySTDinterwordspacing}{\spaceskip=0pt\relax}
\providecommand{\BIBentryALTinterwordstretchfactor}{4}
\providecommand{\BIBentryALTinterwordspacing}{\spaceskip=\fontdimen2\font plus
\BIBentryALTinterwordstretchfactor\fontdimen3\font minus
  \fontdimen4\font\relax}
\providecommand{\BIBforeignlanguage}[2]{{%
\expandafter\ifx\csname l@#1\endcsname\relax
\typeout{** WARNING: IEEEtran.bst: No hyphenation pattern has been}%
\typeout{** loaded for the language `#1'. Using the pattern for}%
\typeout{** the default language instead.}%
\else
\language=\csname l@#1\endcsname
\fi
#2}}
\providecommand{\BIBdecl}{\relax}
\BIBdecl

\bibitem{9064949}
L.~{You}, J.~{Xiong}, A.~{Zappone}, W.~{Wang}, and X.~{Gao}, ``Spectral
  efficiency and energy efficiency tradeoff in massive {MIMO} downlink
  transmission with statistical {CSIT},'' \emph{IEEE Transactions on Signal
  Processing}, vol.~68, pp. 2645--2659, 2020.

\bibitem{al2009much}
T.~Al-Naffouri, M.~Sharif, and B.~Hassibi, ``How much does transmit correlation
  affect the sum-rate scaling of {MIMO} {G}aussian broadcast channels?''
  \emph{IEEE Transactions on Communications}, vol.~57, no.~2, pp. 562--572,
  2009.

\bibitem{4557521}
V.~{Raghavan} and V.~V. {Veeravalli}, ``On quantized multi-user beamforming in
  spatially correlated broadcast channels,'' in \emph{IEEE International
  Symposium on Information Theory}, 2007.

\bibitem{clerckx2008correlated}
B.~Clerckx, G.~Kim, and S.~Kim, ``Correlated fading in broadcast {MIMO}
  channels: Curse or blessing?'' in \emph{IEEE Global Telecommunications
  Conference (GLOBECOM)}, 2008.

\bibitem{clerckx2008mu}
------, ``{MU-MIMO} with channel statistics-based codebooks in spatially
  correlated channels,'' in \emph{IEEE Global Telecommunications Conference
  (GLOBECOM)}, 2008.

\bibitem{huang2010exploiting}
Y.~Huang, L.~Yang, M.~Bengtsson, and B.~Ottersten, ``Exploiting long-term
  channel correlation in limited feedback {SDMA} through channel phase
  codebook,'' \emph{IEEE Transactions on Signal Processing}, vol.~59, no.~3,
  pp. 1217--1228, 2010.

\bibitem{raghavan2013statistical}
V.~Raghavan, S.~V. Hanly, and V.~V. Veeravalli, ``Statistical beamforming on
  the {G}rassmann manifold for the two-user broadcast channel,'' \emph{IEEE
  Transactions on Information Theory}, vol.~59, no.~10, pp. 6464--6489, 2013.

\bibitem{wang2012statistical}
J.~Wang, S.~Jin, X.~Gao, K.-K. Wong, and E.~Au, ``Statistical eigenmode-based
  {SDMA} for two-user downlink,'' \emph{IEEE Transactions on Signal
  Processing}, vol.~60, no.~10, pp. 5371--5383, 2012.

\bibitem{zhang2019robust}
X.~Zhang, J.~Wang, C.~Jiang, C.~Yan, Y.~Ren, and L.~Hanzo, ``Robust beamforming
  for multibeam satellite communication in the face of phase perturbations,''
  \emph{IEEE Transactions on Vehicular Technology}, vol.~68, no.~3, pp.
  3043--3047, 2019.

\bibitem{9110855}
L.~{You}, K.~X. {Li}, J.~{Wang}, X.~{Gao}, X.~G. {Xia}, and B.~{Ottersten},
  ``Massive {MIMO} transmission for {LEO} satellite communications,''
  \emph{IEEE Journal on Selected Areas in Communications}, vol.~38, no.~8, pp.
  1851--1865, 2020.

\bibitem{8019852}
H.~{Joudeh} and B.~{Clerckx}, ``Rate-splitting for max-min fair multigroup
  multicast beamforming in overloaded systems,'' \emph{IEEE Transactions on
  Wireless Communications}, vol.~16, no.~11, pp. 7276--7289, 2017.

\bibitem{8756668}
Y.~{Mao}, B.~{Clerckx}, and V.~O.~K. {Li}, ``Rate-splitting multiple access for
  coordinated multi-point joint transmission,'' in \emph{IEEE International
  Conference on Communications Workshops (ICC Workshops)}, 2019.

\bibitem{mao2018rate}
Y.~Mao, B.~Clerckx, and V.~O. Li, ``Rate-splitting multiple access for downlink
  communication systems: bridging, generalizing, and outperforming {SDMA} and
  {NOMA},'' \emph{EURASIP journal on wireless communications and networking},
  vol. 2018, no.~1, p. 133, 2018.

\bibitem{7434643}
M.~{Dai}, B.~{Clerckx}, D.~{Gesbert}, and G.~{Caire}, ``A rate splitting
  strategy for massive {MIMO} with imperfect {CSIT},'' \emph{IEEE Transactions
  on Wireless Communications}, vol.~15, no.~7, pp. 4611--4624, 2016.

\bibitem{7555358}
H.~{Joudeh} and B.~{Clerckx}, ``Sum-rate maximization for linearly precoded
  downlink multiuser {MISO} systems with partial {CSIT}: A rate-splitting
  approach,'' \emph{IEEE Transactions on Communications}, vol.~64, no.~11, pp.
  4847--4861, 2016.

\bibitem{7513415}
------, ``Robust transmission in downlink multiuser {MISO} systems: A
  rate-splitting approach,'' \emph{IEEE Transactions on Signal Processing},
  vol.~64, no.~23, pp. 6227--6242, 2016.

\bibitem{9257433}
L.~{Yin} and B.~{Clerckx}, ``Rate-splitting multiple access for multigroup
  multicast and multibeam satellite systems,'' \emph{IEEE Transactions on
  Communications}, pp. 1--1, 2020.

\bibitem{8846706}
Y.~{Mao}, B.~{Clerckx}, and V.~O.~K. {Li}, ``Rate-splitting for multi-antenna
  non-orthogonal unicast and multicast transmission: Spectral and energy
  efficiency analysis,'' \emph{IEEE Transactions on Communications}, vol.~67,
  no.~12, pp. 8754--8770, 2019.

\bibitem{8491100}
------, ``Energy efficiency of rate-splitting multiple access, and performance
  benefits over {SDMA} and {NOMA},'' in \emph{2018 15th International Symposium
  on Wireless Communication Systems (ISWCS)}, 2018.

\bibitem{9205852}
Z.~{Lin}, M.~{Lin}, B.~{Champagne}, W.~P. {Zhu}, and N.~{Al-Dhahir}, ``Secure
  and energy efficient transmission for {RSMA}-based cognitive
  satellite-terrestrial networks,'' \emph{IEEE Wireless Communications
  Letters}, pp. 1--1, 2020.

\end{thebibliography}

\end{document}